\documentclass[traditabstract]{aa} 
\usepackage{savesym} 
\usepackage{txfonts} 
\usepackage[english]{babel} 
\usepackage{t1enc} 
\usepackage{graphicx} 
\usepackage{subfig} 
\usepackage{pdflscape}
\usepackage{fixltx2e} 
\usepackage{natbib} 
\usepackage{footmisc}
\usepackage{url} 
\usepackage{longtable}
\usepackage{multirow}
\usepackage{setspace}
\usepackage[table]{xcolor}
\usepackage{gensymb}
\usepackage{siunitx}
\usepackage{adjustbox}
\usepackage{float}
\usepackage{booktabs}
\usepackage{indentfirst}
\usepackage{float}
\usepackage{stfloats}
\usepackage{upgreek}
\usepackage{commath}

\usepackage[hidelinks]{hyperref}

\hypersetup{
  colorlinks   = true, 
  urlcolor     = blue, 
  linkcolor    = blue, 
  citecolor   = blue 
}

\RequirePackage[margin=0cm,font={color=black,small},labelfont={color=black,bf},tableposition=top,singlelinecheck=false,hypcap]{caption} 
\newcommand{\RNum}[1]{\uppercase\expandafter{\romannumeral #1\relax}}

\begin{document} 
 
\title{What you see is not necessarily what you get: Interpreting near-infrared scattering phase functions of debris discs}
\author{Q. Bosschaart\inst{1}, J. Olofsson\inst{1}}

\institute{\inst{1} European Southern Observatory, Karl-Schwarzschild-Strasse 2, 85748 Garching bei München, Germany
\\  e-mail: {\tt Quincy.Bosschaart@eso.org}}

\date{Received ; accepted } 
 
\abstract{\textit{Context.} Scattering phase functions (SPFs) derived from resolved scattered-light images of debris discs are widely used to infer dust grain properties, often through parametric descriptions such as the Henyey-Greenstein (HG) phase function. However, the extent to which the inferred scattering behaviour reflects the intrinsic dust properties, rather than projection effects, disc geometry, or methodological choices, remains uncertain.\\
\textit{Aims.} We investigate how reliably scattering phase functions and HG asymmetry parameters can be recovered from scattered-light images, and identify the regimes in which geometric and methodological effects introduce significant biases.\\
\textit{Methods.} For the first time, we perform a systematic test of the reliability of SPF determinations using a physically motivated forward-modelling framework that combines dust-scattering calculations, size-dependent grain dynamics, and ray-tracing imaging to generate synthetic total-intensity scattered-light observations. Because the intrinsic SPFs are known a priori, surface-brightness phase functions extracted from the images can be directly compared to the input scattering behaviour. We explore a grid of grain size distributions, disc inclinations, and opening angles, and fit two-component Henyey-Greenstein functions to evaluate how reliably the forward-scattering parameter $g_\mathrm{1}$ traces the underlying dust properties.\\
\textit{Results.} Even under idealised conditions with perfect knowledge of the disc geometry, the extracted phase functions can differ substantially from the intrinsic SPFs. First, the limited range of accessible scattering angles plays a dominant role: the smallest angles, where the strongest forward-scattering peaks occur, are generally unobservable. As a result, the apparent scattering behaviour does not vary monotonically with grain size: large grains may not appear as forward-scattering, while grains comparable to the observing wavelength can appear comparatively more anisotropic. Second, projection effects, line-of-sight mixing, and SPF-extraction choices further modify the recovered phase functions. Consequently, the fitted HG asymmetry parameter $g_\mathrm{1}$ depends strongly on viewing geometry and methodology and does not uniquely trace grain properties.\\
\textit{Conclusions.} Scattering phase functions and HG asymmetry parameters derived from scattered-light images should be interpreted as effective, observation-dependent quantities rather than direct proxies of dust properties. Robust interpretation of debris-disc images therefore requires forward modelling that accounts for projection effects, limited scattering-angle coverage, and observational biases.

}

\keywords{} 

\titlerunning{Interpreting near-infrared scattering phase functions of debris discs}
\maketitle 
 
\section{Introduction}  
\label{sec:Intro}
\noindent Debris discs are composed of dust grains continuously replenished by collisions of planetesimals and smaller bodies, and are widely observed around main-sequence stars across a broad range of stellar masses (e.g. \citealt{Wyatt2008}; \citealt{Hughes2018}). As such, they offer key insights into the architecture, composition, and long-term evolution of planetary systems, as well as provide direct constraints on the presence of planets shaping the dust distribution (see e.g. \citealt{Mouillet1997}; \citealt{Lagrange2025}). Because the micron-sized dust grains are short-lived compared to stellar ages, their presence reflect an ongoing balance between collisional production and removal processes such as radiation pressure and stellar winds (\citealt{Burns1979}; \citealt{Krivov2010}).\\
\indent In scattered light, debris disc observations offer a direct probe of the small dust grain population. At optical and near-infrared wavelengths, the contribution to the surface brightness (SB) is dominated by micron-sized particles, whose scattering efficiencies peak when their sizes are comparable to the observed wavelength. The brightness and morphology of the scattered-light images depend strongly on the physical properties of the grains, such as the grain size, composition, porosities and shapes (e.g. \citealt{Kirchschlager_Wolf2014}; \citealt{Hughes2018}; \citealt{Pawellek2024}). Extracting this information from observations relies on understanding how dust grains scatter stellar light, which is commonly described by the scattering phase function (SPF). The SPF characterises the angular dependence of scattering, encoding how efficiently grains scatter light as a function of the scattering angle between the incident stellar radiation and the observer's line of sight.\\
\indent Empirical constraints on scattering phase functions are also available from Solar System dust populations, such as zodiacal dust and cometary grains. Observations of these populations reveal complex phase functions that often exhibit a pronounced forward-scattering peak, a relatively flat behaviour at intermediate scattering angles, and, in some cases, enhanced back-scattering. Such behaviour is generally attributed to dust composed of irregular, highly porous aggregates rich in organic material (e.g. \citealt{Levasseur-Regourd2020}). SPFs inferred for debris discs from scattered-light imaging exhibit similar qualitative features (e.g. \citealt{Graham2007}; \citealt{Hughes2018}; \citealt{Hom2024}), despite the very different stellar environments, suggesting that Solar System dust populations may provide a useful analogue for debris disc grains.\\
\indent Regardless of these qualitative similarities, the interpretation of SPFs in debris discs ultimately depends on how scattering behaviour is linked to the physical properties of the dust. In particular, the shape of the SPF is strongly linked to dust physics. Small grains tend to scatter light more isotropically, whereas larger grains generally exhibit increasingly anisotropic scattering in total intensity, with a stronger preference for forward-scattering at small scattering angles (see \citealt{Mulders2013}). As a result, the surface brightness distribution observed in spatially resolved scattered-light images can ultimately provide constraints on the physical properties of dust in debris discs, such as grain size distributions and scattering behaviour (e.g. \citealt{Olofsson2016}; \citealt{Milli2017}). In practice, however, the interpretation of SPFs derived from resolved scattered-light images is far from straightforward. In debris discs, radiation pressure strongly affects the dynamics of small grains, producing size-dependent spatial distributions in which the grain size distribution varies with distance from the star (e.g. \citealt{Strubbe_Chiang2006}; \citealt{Thebault_Wu2008}). As a consequence, the surface brightness measured at any given location represents the integrated scattering signal from a range of grain sizes with different optical properties, rather than a single, well-defined population. This spatial and size-dependent mixing of scattering properties is generally not accounted for in simplified analyses, which often assume uniform dust properties throughout the disc (e.g. \citealt{Olofsson2020}). In addition, scattered-light observations provide a two-dimensional projection of a three-dimensional dust distribution, and only a limited range of scattering angles can be accessed, depending on the disc inclination, radial extent, and vertical thickness (e.g. \citealt{Stark2014}; \citealt{Milli2017}; \citealt{Milli2019}). Despite these limitations, a common approach in observational studies is to extract an effective SPF from the observed surface brightness distribution and to describe it using simple parametric phase functions, most notably the Henyey-Greenstein (HG) phase function (\citealt{Henyey_Greenstein1941}) or combinations thereof (e.g. \citealt{Schneider2006}; \citealt{Debes2008}; \citealt{Thalmann2013}, or more recently \citealt{Milli2019}; \citealt{Engler2020}; \citealt{Desgrange2025}). The resulting HG asymmetry parameter is frequently interpreted as a proxy for grain size or overall scattering anisotropy, and is often compared between different systems (e.g. \citealt{Hughes2018}; \citealt{Engler2020}). However, the extent to which such inferred parameters reflect the true dust scattering physics, rather than the effects of viewing geometry, projection, or the adopted SPF-extraction method, is not entirely straightforward.\\
\indent A quantitative assessment of these effects requires a controlled framework in which the underlying scattering phase function is known a priori, and the effects of disc geometry, viewing inclination, and analysis methodology can be isolated. Forward-modelling approaches provide such a framework by generating synthetic scattered-light images from physically motivated dust models, allowing the same phase-function extraction techniques used on observations to be applied in a fully controlled setting.\\
\indent In this paper, we use a forward-modelling framework to systematically test these effects. Using synthetic observations, we compare the extracted SPFs to the known intrinsic SPFs, and examine how commonly adopted HG parametrisations respond to these effects. This approach allows us to assess the reliability and limitations of inferred asymmetry parameters across a wide range of dust properties and disc configurations, and to identify the regimes in which geometric and methodological effects are most significant.\\
\indent The paper is organised as follows. In Section \ref{sec:modelling_framework_bound}, we describe the dust scattering models, the prescription for dust dynamics, and the ray-tracing framework used to generate synthetic scattered-light images. Section \ref{sec:test_pipeline} details the methodology for extracting surface-brightness phase functions from these synthetic images. In Section \ref{sec:reliability}, we assess the reliability of the extracted SPFs and quantify how observational effects, disc geometry, and methodological choices influence the recovered scattering behaviour. Section \ref{sec:HG} examines how these effects propagate into fitted HG asymmetry parameters. Finally, in Section \ref{sec:discussion}, we discuss the implications of our results for interpreting scattered-light observations of debris discs, and we summarise our findings in Section \ref{sec:conclusion}.

\section{Modelling framework}
\label{sec:modelling_framework_bound}
\noindent In this section, we describe the forward-modelling framework used to generate the synthetic scattered-light images analysed in this work. The framework combines dust-scattering calculations, a prescription for the size-dependent spatial distribution of grains driven by radiation pressure and orbital dynamics, and ray-tracing imaging to produce resolved disc images for a range of grain populations and disc geometries.
\subsection{Dust scattering properties with \texttt{optool}}
\noindent The scattering properties of dust grains in debris discs are computed using \texttt{optool} (\citealt{optool}), which generates SPFs for a given set of grain parameters. To approximate the irregular shapes of real dust particles, we adopt the Distribution of Hollow Spheres (DHS) method (\citealt{Min2005}), which improves on classical Mie theory by capturing the effects of non-sphericity. Throughout this study, we keep the dust composition fixed to avoid introducing additional degeneracies. The composition corresponds to \texttt{"pyr-mg70 0.87 c 0.13 -m h2o-w 0.2 -p 0.5"}, consistent with a mixture of magnesium-rich pyroxene silicate (Mg$_\mathrm{0.7}$Fe$_\mathrm{0.3}$SiO$_\mathrm{3}$, \citealt{Dorschner1995}), which constitutes 87\% of the solid volume, and amorphous carbon (\citealt{Zubko1996}, 13\%). To account for the possibility of icy parent-body material in the outer regions, we include a 20\% water-ice fraction as a mantle coating the grains (e.g. \citealt{Donaldson2013}). The grains are assumed to be highly porous, with 50\% void volume, reflecting the fluffy structure expected for fragments produced in high-velocity planetesimal collisions (see e.g. \citealt{Augereau1999}). Such mixtures of silicate, carbonaceous material, ice mantles, and porosity are widely used in debris disc modelling (see e.g. the review of \citealt{Hughes2018}). Fixing the composition reduces the dimensionality and allows us to focus on the effects of other parameters on the resulting SPFs.\\
\indent The grain sizes follow a power-law distribution, $\text{d}N/\text{d}s \propto s^{q}$, with $s$ the grain size and $q$ the power-law index. In our models, the minimum grain size, $s_\mathrm{min}$, and the power-law index are varied, while the maximum grain size is fixed at 1000 $\upmu$m, several orders of magnitude larger than the observing wavelength. We explore values of $s_\mathrm{min}$ between 0.01 and 100 $\upmu$m, encompassing the range typically inferred from modified blackbody fits and from mid-infrared silicate feature analysis, which often indicate minimum grain sizes of a few to a few tens of microns (\citealt{Holland2017}; \citealt{Sibthorpe2018}; \citealt{Mittal2015}). Extending the grid values to below 1 $\upmu$m and above several tens of microns allows us to probe extreme regimes. The power-law index is varied between -3 and -4, consistent with slopes expected from steady-state collisional models (see e.g. \citealt{Pan_Sari2005}; \citealt{Gaspar2012}; \citealt{Pan_Schlichting2012}).

\subsection{Ray-tracing imaging with \texttt{bound}}
\noindent To compute synthetic scattered-light observations, we use \texttt{bound}, an extended and adapted version of the \texttt{betadisk} framework described by \citet{Olofsson2022}. Similarly, \texttt{bound} uses a radiation pressure based parametrisation to place dust grains on their orbits analytically, avoiding the need for explicit orbit integration. This makes the method extremely fast and ideal for modelling debris disc observations. However, with this approach we cannot account for the contribution of unbound grains (e.g. \citealt{Thebault_Kral2019}).\\
\indent \texttt{bound} begins by defining the orbital architecture of the birth ring, the region where parent bodies (planetesimals) continuously collide and release dust. The orbits of the parent bodies are specified by their semi-major axis, eccentricity, and inclination relative to the disc midplane. In this study, we assume zero eccentricity ($e = 0$) for simplicity, so that all dynamical effects originate purely from grain-size-dependent radiation pressure. The vertical distribution of dust is characterised by an effective opening angle, defined as the typical inclination of dust grains relative to the midplane.\\
\indent Dust grains are drawn from the size distribution and for each grain, the radiation pressure coefficient $\beta$ is computed. While \texttt{bound} is capable of calculating a "proper" $\beta$ from the full optical properties (composition, porosity, refractive indices) and stellar parameters, in this work we intentionally adopt the simplified size-scaling relation $\beta = 0.5s_\mathrm{min}/s$, so that the minimum grain size is equal to the blow-out size. This avoids mixing the sensitivity of the SPF to grain properties with the dynamical redistribution of the grains, allowing us to examine how the scattering behaviour inferred from synthetic images compares to the intrinsic expectations and to quantify the level of confidence that can be placed in retrieved phase functions. With $\beta$ known, each dust particle immediately acquires an eccentricity approximately equal to $\beta  /  (1  - \beta$), reflecting the effect of radiation pressure (e.g \citealt{Wyatt1999}; \citealt{Wyatt2006}), and its orbit is analytically modified accordingly. Each particle is then placed at a random location on this orbit, uniformly in mean anomaly to account for Kepler's second law, and projected onto the sky using a given inclination and position angle of the disc.\\
\indent Once projected onto the image plane, each grain contributes to the local brightness according to its scattering angle $\theta$, the angle between the incoming starlight and the direction toward the observer. The total-intensity scattering signal is set by the \texttt{optool}-derived Müller matrix element $S_\mathrm{11}$, the scattering efficiency $Q_\mathrm{sca}$, and the grain cross section. For a grain of size $s$ the scattered intensity scales as $I(\theta)$ $\propto$ $S_\mathrm{11}(\theta) \ Q_\mathrm{sca} \ \pi s^\mathrm{2}$ / ($4\pi r^\mathrm{2}$). Grain sizes are sampled according to the size distribution, with weights applied to account for the fact that fewer collisions occur in the disc halo (\citealt{Strubbe_Chiang2006}; \citealt{Thebault_Wu2008}). In this study, we only use the total-intensity scattered-light images, but \texttt{bound} is also capable of producing polarised scattered-light images (see Section \ref{sec:discussion_polarisation}) and thermal emission maps (e.g. for modelling ALMA (Atacama Large Millimeter/submillimeter Array) observations).
\subsubsection{Model setup}
\noindent We generate total-intensity scattered-light images with $10^{6}$ particles, using the fixed dust composition described above. Models are computed at a single wavelength, $\lambda = 1.63$ $\upmu$m, corresponding to the SPHERE (Spectro-Polarimetric High-contrast Exoplanet REsearch instrument at the VLT) H-band central wavelength. While \texttt{bound} can produce images at arbitrary wavelengths, integrating across a filter transmission curve increases computation time. The effect of the full SPHERE H-band throughput is explored in Section \ref{sec:wavelength}.\\
\indent For our main analysis, we vary $s_\mathrm{min}$ and $q$ to capture the effect of dust size distribution on the scattering phase function. We also vary the inclination and opening angle to assess how viewing geometry and vertical structure influence the observable signal. These parameters are chosen because they most strongly impact the measured SPF, either by changing the relative contributions of different grain sizes or by modifying the range of accessible scattering angles. This approach allows us to quantify how SPFs extracted from synthetic images deviate from the true \texttt{optool} input phase function and how such deviations affect standard dust analysis techniques.
\subsubsection{A first look: grain size and apparent forward scattering}
\label{sec:synthetic_images}
\begin{figure*}[ht]
    \includegraphics[scale=0.44]{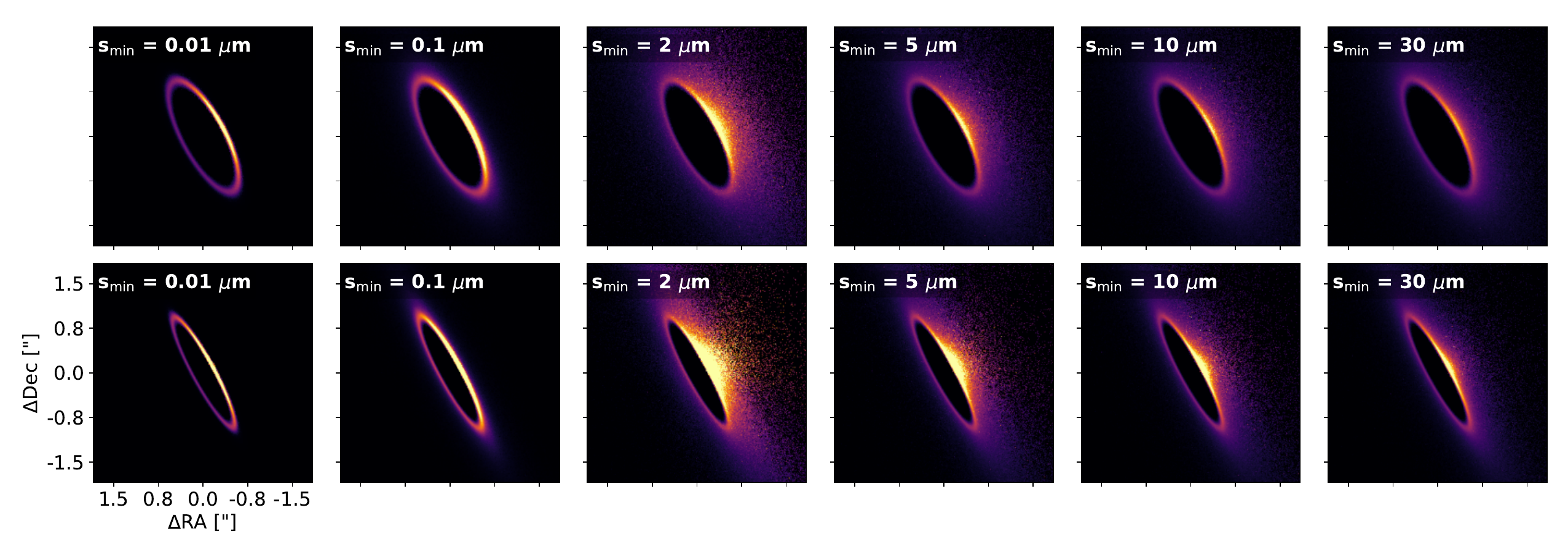}
    \caption{Gallery of synthetic total-intensity scattered-light images generated with \texttt{bound}. The opening angle and power-law index are fixed, while the minimum grain size increases from left to right. The two rows correspond to different disc inclinations: $i = 70^{\circ}$ for the first row and $i = 80^{\circ}$ for the second row. All images are shown with the same power-law colour scale.}
\label{fig:gallery_bound}
\end{figure*}
\noindent Before analysing how scattering phase functions are extracted from scattered-light images, it is instructive to examine what the synthetic \texttt{bound} images themselves look like and how their appearance changes with dust properties and viewing geometry. This step is essential because scattered-light images are often interpreted qualitatively, and such visual interpretations implicitly assume a direct correspondence between apparent brightness asymmetries and the underlying dust scattering phase function. Figure \ref{fig:gallery_bound} presents a gallery of total-intensity scattered-light images for a representative subset of our models. The opening angle and $q$ are fixed at 0.01 rad and -3.5, respectively, while the minimum grain size increases from left to right. The two rows correspond to different inclinations, with the top row showing a moderately inclined disc (70$^{\circ}$) and the bottom row a more highly inclined configuration (80$^{\circ}$).\\
\indent At first glance, the apparent scattering behaviour inferred from these images is counter-intuitive. From dust-scattering theory (see e.g. \citealt{Hughes2018}), one would expect discs dominated by very small grains ($s_\mathrm{min} < \lambda$) to exhibit nearly isotropic scattering and therefore a relatively symmetric brightness distribution across the disc. As the grain size increases, scattering should become progressively more forward-peaked, leading to an increasingly strong brightness asymmetry between the near and far sides of the disc. The synthetic images, however, do not follow this simple trend. Discs dominated by the smallest grains already show a noticeable brightness asymmetry, indicating more forward scattering than expected for nearly isotropic SPFs. Models with intermediate minimum grain sizes ($s_\mathrm{min} \sim \lambda$) exhibit the strongest apparent forward scattering. For even larger grains ($s_\mathrm{min} = 10$ or 30 $\upmu$m), the discs instead appear nearly isotropic.\\
\indent These synthetic images demonstrate that the apparent scattering behaviour seen in scattered-light images does not directly reflect the underlying dust scattering phase function. This discrepancy arises because scattered-light images sample only a restricted range of scattering angles, set by the disc inclination and geometry, rather than the full intrinsic phase function. As a result, strongly forward-scattering grains can appear comparatively isotropic if their most anisotropic scattering occurs at angles that are not observable. Moreover, even in this controlled set of models, it is not always straightforward to quantify the degree of forward scattering from visual inspection alone. For example, differences in brightness asymmetry between models can be subtle, and it is not immediately clear how such variations should be interpreted in terms of the underlying phase function.\\
\indent This motivates a more quantitive assessment of how reliably SPFs can be extracted from scattered-light images, and how confidently the resulting phase functions can be interpreted in terms of dust properties. In particular, the counter-intuitive trends seen in Figure \ref{fig:gallery_bound} highlight that apparent brightness asymmetries alone are not sufficient to infer the underlying dust scattering behaviour. Because the intrinsic SPFs are known in our simulations, the extracted phase functions can be directly compared to the input, providing a reference against which the reliability of the measurement can be evaluated. In the following section, we therefore investigate how SPFs derived using a widely adopted direct extraction method compare to the true phase functions, and how observational and modelling choices influence the inferred scattering behaviour.

\section{Methodology: determination of the SPF}
\label{sec:test_pipeline}
\begin{figure}[ht]
    \includegraphics[scale=0.43]{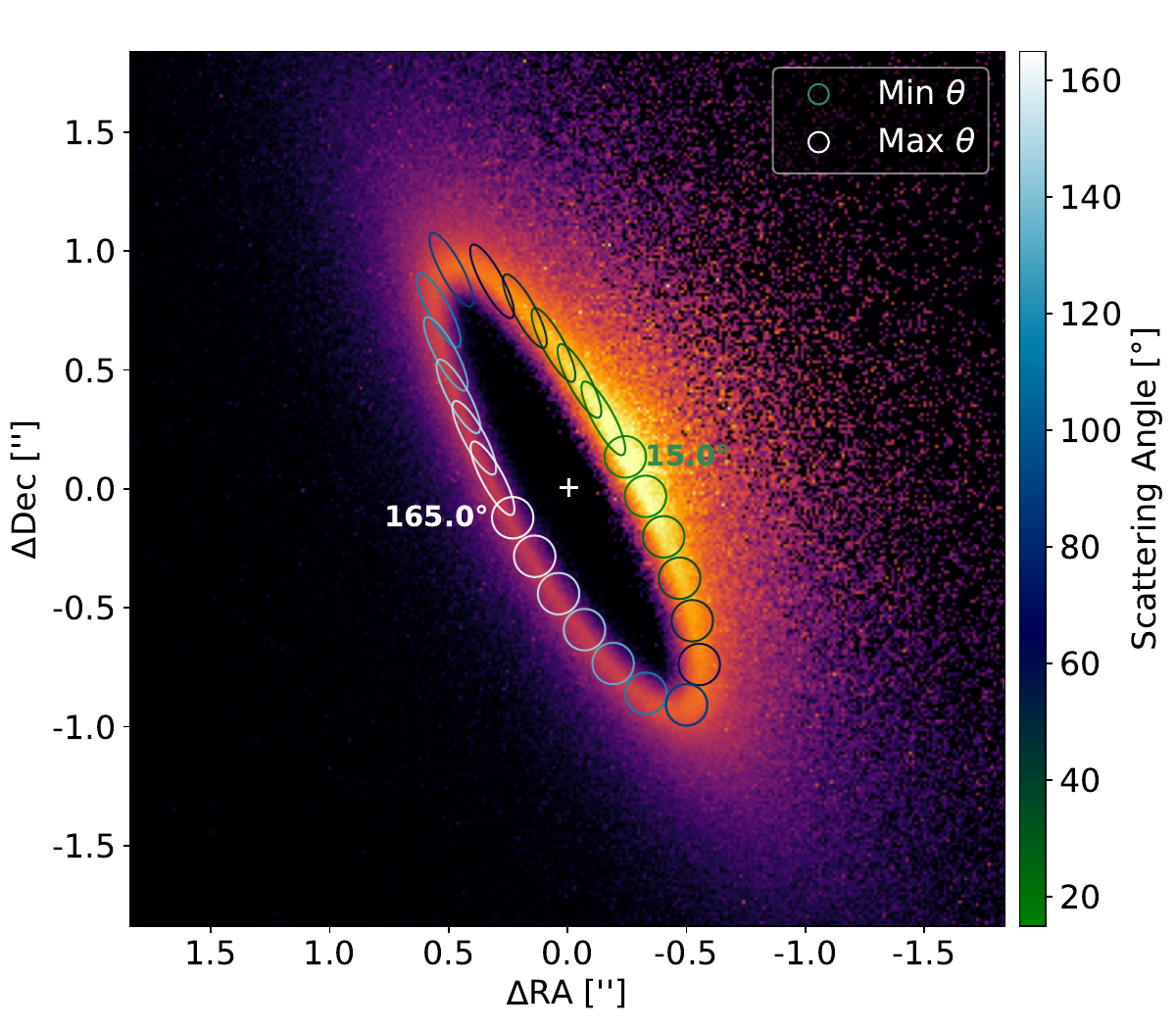}
    \caption{Log-scaled synthetic image of a \texttt{bound} debris disc, highlighting both the birth ring and the extended halo. The SB sampling scheme is overplotted: elliptical apertures on one half of the disc account for disc inclination, while circular apertures on the other half are shown for comparison. Apertures are colour-coded by scattering angle, with the minimum and maximum scattering angles indicated. RA and Dec offsets from the disc centre are shown in arcseconds on the x- and y-axes, respectively.}
    \label{fig:ellipse_plot}
\end{figure}
\subsection{The standard direct-extraction method}
\label{sec:standard_method}
\noindent Several techniques have been used in the literature to extract SPFs from scattered-light images. One of the main approaches is a direct extraction method, where the surface brightness is measured at different scattering angles along the disc (e.g. \citealt{Milli2017, Milli2019}; \citealt{Engler2019}; \citealt{Ren2019}). Another technique relies on forward modelling, in which a parametric SPF is adopted and the disc image is generated, convolved with the instrumental point spread function (PSF), and compared to the data to optimise the underlying parameters (e.g \citealt{Chen2020}; \citealt{Mazoyer2020}). A third, iterative non-parametric approach has been proposed for polarimetric data, where an estimate of the dust density distribution is compared to the data to derive the phase function best suited to match the observations (\citealt{Olofsson2020}). \citet{Milli2024} combined the two main approaches for a disc with a small inclination ($\sim$30$^{\circ}$) and found that the two methods yielded mutually consistent SPFs within the uncertainties. Therefore, in this work, we adopt the direct-extraction technique, building on the formalism of \citet{Milli2017}. Because the synthetic images analysed here are generated with \texttt{bound}, the orbital architecture of the birth ring is known exactly, allowing us to isolate methodological effects without uncertainties in the disc geometry.\\
\indent This method relies on the fact that, for a geometrically thin ring each sky-projected position along the ring corresponds to a unique scattering angle, thereby adopting two assumptions: (i) the scale-height is assumed to be small compared to the ring radius, ensuring a one-to-one mapping between sky position and scattering angle; and (ii) variations in surface brightness along the ring are interpreted as arising solely from the angular dependence of the SPF, rather than from intrinsic changes in dust density or composition.\\
\indent Given the semi-major axis $a_\mathrm{0}$ of the birth ring, the inclination $i$, and the position angle (PA) of the ring, the birth ring projects onto the sky as an ellipse with semi-axes $a_\mathrm{sky} = a_\mathrm{0}$ and $b_\mathrm{sky} = a_\mathrm{0}\cos$($i$). We sample this ellipse at regular intervals in arc length, placing sampling points uniformly along the projected ring. The spacing between neighbouring points is chosen to be a multiple of the PSF full width at half maximum (FWHM). This ensures that the surface-brightness measurements performed at each location probe largely independent regions of the image, while still sampling the ring densely enough to trace the variation of scattering angle. In practice, we adopt the FWHM of a representative SPHERE H-band PSF, 0.044", which also enables us to create PSF-convolved versions of the \texttt{bound} models for a realistic comparison with observed data (see Section \ref{sec:convolution}).\\
\indent For each point ($x$, $y$) on the sky plane, the corresponding three-dimensional position in the disc plane is obtained by deprojecting the ellipse, assigning a vertical coordinate $z$ = $y$ $\tan$($i$), such that the resulting vector ($x$, $y$, $z$) points from the star to the sampled location in the disc. The scattering angle $\theta$ is then defined as the angle between the vector and the line of sight. The range of scattering angles probed is
\[
\qquad \qquad \theta_\mathrm{min} = 90^{\circ} - i, \qquad \qquad \theta_\mathrm{max} = 90^{\circ} + i, \]
with forward scattering on the near side of the disc (small values of $\theta$) and backward scattering on the far side (large values of $\theta$). We apply this sampling separately to both sides of the minor axis of the ring, ensuring full coverage of the scattering-angle domain.\\
\indent At each sampled location we measure the local surface brightness using aperture photometry. Because the disc is inclined, circular apertures in the disc plane project onto the sky as ellipses. In studies that explicitly derive scattering phase functions, the sampling of the projected birth ring generally accounts for this geometric effect to preserve the one-to-one relation between sky position and scattering angle (e.g. \citealt{Milli2017}; \citealt{Engler2019}). In contrast, SB profiles in debris discs are sometimes measured using circular or rectangular apertures in the image plane (e.g. \citealt{Debes2009}). In Section \ref{sec:ellipses}, we assess the impact of these different aperture definitions on the extracted surface brightness. The aperture radius is set to a fraction of the PSF FWHM, maximising the signal enclosed while reducing contamination from other scattering angles. The sum of the enclosed pixel values within each aperture is then normalised by the inverse square of the physical distance from the star to each sampled point, to account for the $1/r^\mathrm{2}$ dependence of the incident stellar flux. The resulting quantity represents the flux per scattering angle. The measurements from the northern and southern sides of the disc are combined to provide an estimate of the SPF. Figure \ref{fig:ellipse_plot} shows an example of the sampling scheme for a \texttt{bound} model viewed at $i = 75^{\circ}$ and PA$ = -151.6^{\circ}$, values chosen to be similar to those of the well-studied debris disc HR 4796 (\citealt{Olofsson2022_HR4796}). The parameters chosen for this specific model are $s_\mathrm{min} = 10$ $\upmu$m, $q = -3.5$, opening angle $= 0.01$ rad, and $a_\mathrm{0} = 1.04$". The near side of the disc corresponds to the smallest scattering angles, denoted by $\theta_\mathrm{min}$, while the far side corresponds to the largest scattering angles, denoted by $\theta_\mathrm{max}$. For illustration, elliptical apertures are shown on one side of the disc, with circular apertures displayed on the other.\\

\subsection{Quantifying differences between phase functions}
\label{sec:diff_SPF_SB}
\noindent To enable a systematic comparison between different phase functions derived throughout this work, we require a quantitative metric that captures differences in shape over the accessible scattering angles. We therefore define a scalar measure that quantifies the integrated absolute difference between two normalised phase functions. Given two phase functions $f_\mathrm{1}(\theta)$ and $f_\mathrm{2}(\theta)$, both normalised at $\theta = 90^{\circ}$, we define 
\begin{equation}
    \Delta Area = \int^\mathrm{\theta_\mathrm{max}}_\mathrm{\theta_\mathrm{min}} \abs{f_\mathrm{1}(\theta) - f_\mathrm{2}(\theta)} d\theta
\label{eq:delta_area}
\end{equation}
This metric yields a single positive value that reflects how strongly two curves differ in shape over the sampled scattering-angle range. In the section that follows, this metric is used to compare surface-brightness phase functions obtained under different methodological choices, observational assumptions, and physical disc parameters.

\section{Reliability of SPF measurements}
\label{sec:reliability}
\noindent Even under idealised conditions with perfect knowledge of the disc geometry and noise-free images, the extraction of scattering phase functions from resolved images is subject to several sources of bias. In this section, we assess how different methodological and observational choices affect the recovered surface-brightness phase functions, and to what extent these effects can be controlled.

\subsection{Sensitivity to aperture definition}
\label{sec:ellipses}
\begin{figure}[ht]
    \includegraphics[scale=0.26]{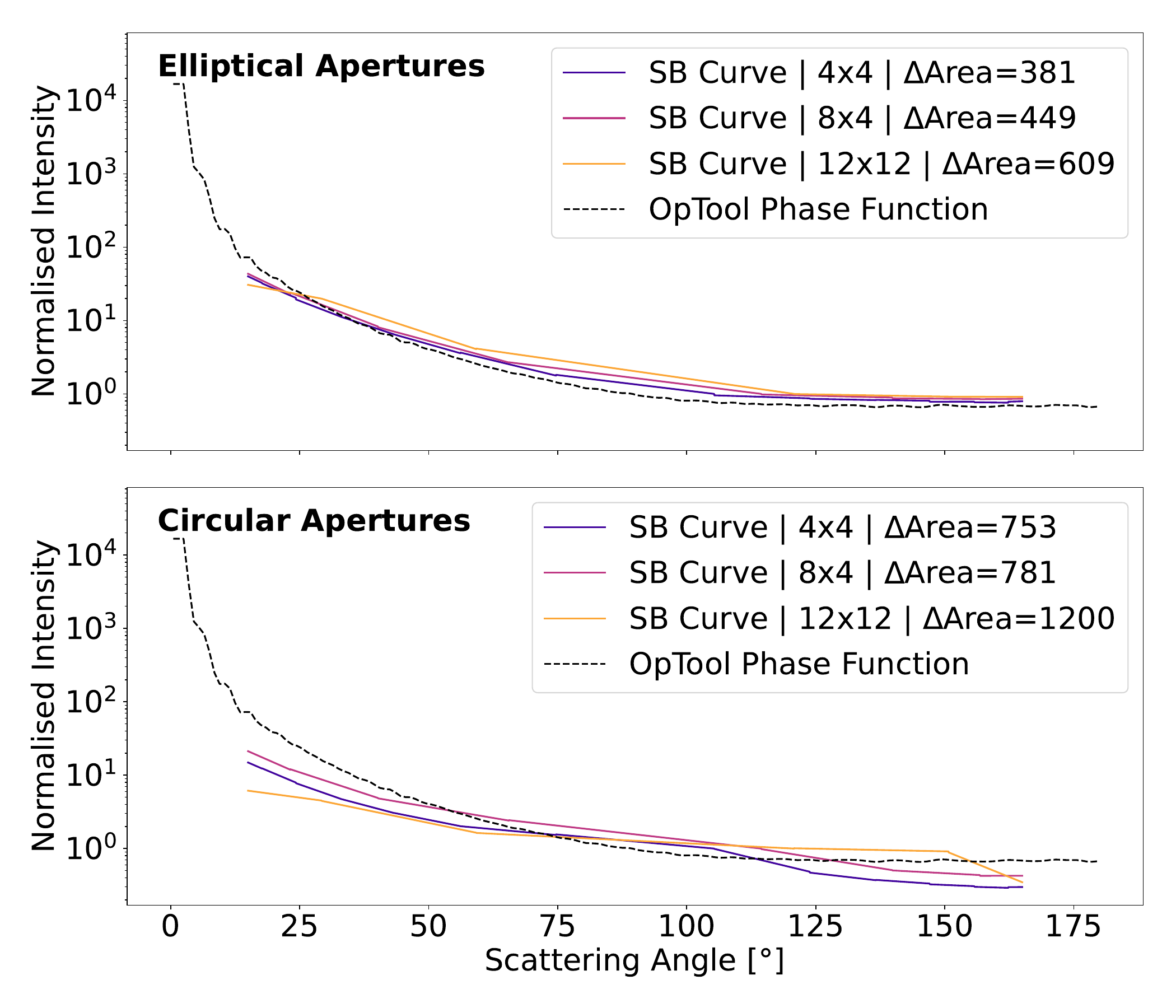}
    \caption{Impact of aperture choice on the extracted SB for a fixed synthetic disc model. Upper panel: SB curves obtained using three different elliptical apertures, compared to the intrinsic SPF. Lower panel: SB curves obtained using three different circular apertures. Each curve is normalised at 90$^{\circ}$, and the legend indicates the sampling and aperture size in units of the PSF FWHM (first number and second number, respectively) along with the integrated absolute difference ($\Delta$Area; Eq. \ref{eq:delta_area}).}
\label{fig:ellipse_vs_circles}
\end{figure}
\noindent We apply different aperture definitions to the same synthetic scattered-light image, keeping all dust parameters and geometric properties fixed. This allows us to assess how strongly the measured SB depends on the choice of aperture alone.\\
\indent Figure \ref{fig:ellipse_vs_circles} compares the extracted SB curves with the corresponding input SPF for three different elliptical aperture configurations (upper panel) and three different circular aperture configurations (lower panel). For each aperture, the resulting $\Delta$Area (Eq. \ref{eq:delta_area}) is also shown. Because the underlying dust parameters and disc geometry are identical for all curves, any differences between the extracted SB curves arise solely from the aperture choices used during the extraction.\\
\indent Changing the aperture size or sampling space modifies how flux is averaged along the ring. Smaller apertures and finer sampling preserve sharper variations in surface brightness but only sample tightly around the birth ring of the disc, whereas larger apertures capture more of the elliptical orbits but where the flux is averaged over a wider range of scattering angles. This geometric averaging alone is sufficient to alter the recovered SB shape, demonstrating that there is no uniquely defined way to extract the SB even under idealised conditions.\\
\indent The effect is considerably more pronounced when circular apertures are used. Because circular apertures do not account for the projection of the inclined disc, they collect light from a broader range of physical locations and scattering angles than elliptical apertures at the same sky position. This leads to stronger distortions in the extracted SB curves and systematically larger $\Delta$Area values compared to elliptical apertures. The enhancement is particularly clear around scattering angles close to 90$^{\circ}$. At these angles, projection effects lead to geometric mixing that dominates the extracted surface brightness.\\
\indent These tests demonstrate that the extraction of the SB from scattered-light images is inherently non-unique and sensitive to methodological choices, even before accounting for dust physics or disc geometry. Because elliptical apertures consistently yield smaller deviations and better preserve the mapping between sky position and scattering angle, we adopt the inclination-corrected elliptical apertures for the remainder of this study.

\subsection{Single wavelength vs full H-band integration}
\label{sec:wavelength}
\begin{figure*}[ht]
    \includegraphics[scale=0.24]{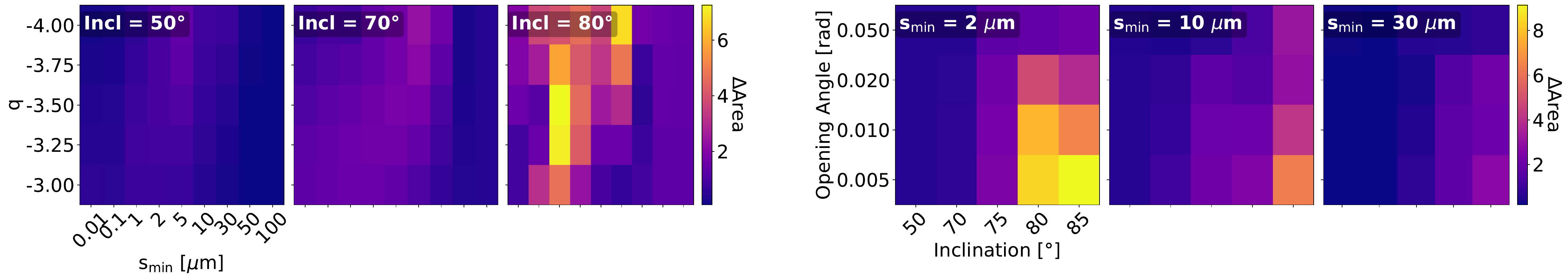}
    \caption{Heatmaps showing the integrated absolute difference ($\Delta$Area; Eq. \ref{eq:delta_area}) between the extracted SB for models at a fixed wavelength of 1.63 $\upmu$m and for models using the full SPHERE H-band transmission profile. The left panels show variations with size distribution slope, $q$, and minimum grain size, $s_\mathrm{min}$, at fixed geometry, for inclinations of 50$^{\circ}$, 70$^{\circ}$, and 80$^{\circ}$. The right panels display the variations with disc opening angle and inclination for $s_\mathrm{min}$ = 2, 10, and 30 micron.}
\label{fig:diff_H_and_lambda}
\end{figure*}
\noindent To test whether adopting a single representative wavelength introduces a bias, we compared scattered-light images computed at a fixed wavelength of 1.63 $\upmu$m with images generated by integrating over the full SPHERE H-band transmission profile. The filter transmission curve was obtained from the SVO Filter Profile Service, which provides instrument throughput curves as a function of wavelength (\citealt{Rodrigo2024}; \citealt{Rodrigo2012}; \citealt{Rodrigo2020}). The transmission curve was sampled using a set of wavelength bins. A synthetic image $I_\mathrm{\lambda}(x, y)$ was computed at each wavelength and the resulting images were combined using transmission-weighted averaging appropriate for an energy-counting detector. The band-integrated intensity was computed as
\begin{equation}
    I_\mathrm{band}(x, y) = \frac{\int T(\lambda) \ \lambda \ I_\mathrm{\lambda}(x, y) \ d\lambda}{\int T(\lambda) \ \lambda \ d\lambda} \ ,
\label{eq:full_H_band}
\end{equation}
where $T(\lambda)$ is the filter transmission. Surface brightness curves were then extracted from both the single-wavelength and broadband images and normalised at a scattering angle of 90$^{\circ}$.\\
\indent To evaluate how this difference depends on both dust properties and disc geometry, we performed this comparison across a grid of models that we will use throughout the rest of this study. In a first set of models, we varied the grain size distribution by spanning $s_\mathrm{min}$ from 0.01 to 100 $\upmu$m and $q$ from -3 to -4, while keeping the disc geometry fixed with a small opening angle of 0.01 rad in order to minimise additional limb-brightening effects. This analysis was carried out for inclinations of 50$^{\circ}$, 70$^{\circ}$, and 80$^{\circ}$. In a second set of models, we varied the opening angle between 0.005 and 0.05 rad for inclinations of 50$^{\circ}$ to 85$^{\circ}$, this time keeping $q$ fixed at -3.5 and adopting representative minimum grain sizes of 2, 10, and 30 $\upmu$m.\\
\indent To visualise how the differences between the extracted SB curves depend on these parameters, we summarise the resulting $\Delta$Area (Eq. \ref{eq:delta_area}) in a series of heatmaps (Figure \ref{fig:diff_H_and_lambda}). Each heatmap encodes the integrated absolute difference as a colour scale, allowing parameter combinations that produce the largest deviations to be identified. The left panels show variations with $q$ and $s_\mathrm{min}$ at fixed geometry, while the right panels show variations with opening angle and inclination.\\
\indent Across the entire parameter space, the resulting $\Delta$Area values are very small, indicating that the extracted SB curves are nearly identical. Slight differences occur for models with small minimum grain sizes (below approximately 10 micron) and at higher inclinations combined with small opening angles. These combinations correspond to cases where the scattering phase function varies more rapidly with wavelengths, such that integrating over the H-band bandwidth introduces some smoothing of the extracted SB. At higher inclinations, projection effects enhance sensitivity to differences between the two methods, and for small opening angles the disc is geometrically thin, so the ellipse sampling closely traces the birth ring and enhances these subtle variations. In contrast, for large opening angles, much of the scattered light originates above or below the birth ring, so a significant fraction falls outside the apertures tracing the ring, making the extracted SB less sensitive to wavelength-dependent differences.\\
\indent This shows that the computationally expensive integration over the full filter transmission curve does not significantly alter the extracted SB compared to adopting a single representative wavelength. We therefore adopt a single wavelength for the remainder of this work, which significantly reduces computation time while preserving the relevant scattering behaviour.

\subsection{Effect of PSF convolution}
\label{sec:convolution}
\begin{figure*}[ht]
    \includegraphics[scale=0.24]{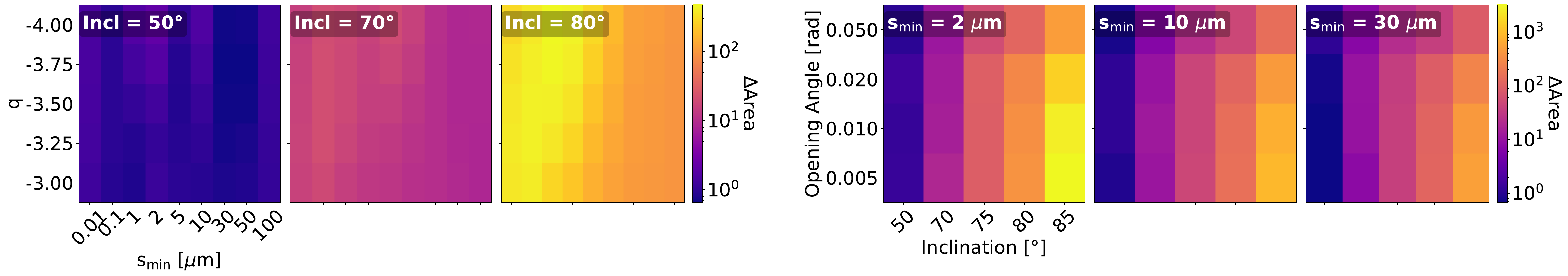}
    \caption{Heatmaps showing the integrated absolute difference ($\Delta$Area; Eq. \ref{eq:delta_area}) between the convolved and unconvolved SB curves. The left panels show variations with size distribution slope, $q$, and minimum grain size, $s_\mathrm{min}$, at fixed geometry, for inclinations of 50$^{\circ}$, 70$^{\circ}$, and 80$^{\circ}$. The right panels display the variations with disc opening angle and inclination for $s_\mathrm{min}$ = 2, 10, and 30 micron.}
\label{fig:diff_conv_and_unconv}
\end{figure*}
\noindent To assess how realistic observational effects influence the extracted surface brightness curves, we convolve the synthetic scattered-light images with a representative SPHERE H-band PSF. In real observations, the disc emission is always blurred by the instrumental PSF, which smooths spatial structure and mixes signals from neighbouring pixels. Applying PSF convolution to our models therefore allows us to quantify how this smoothing alone modifies the SB curves extracted with the ellipse-based method. We model the PSF as a normalised two-dimensional Gaussian with a FWHM matching that measured from the representative SPHERE observation (0.044"). The convolution is applied directly to the synthetic image, after which the SB extraction is performed identically on both the convolved and unconvolved images. Again, the $\Delta$Area metric from Eq. \ref{eq:delta_area} is used to quantify the difference between the two extracted SB curves. The resulting $\Delta$Area values are shown as heatmaps in Figure \ref{fig:diff_conv_and_unconv}.\\
\indent From the heatmaps it is clear that the values of $\Delta$Area are substantially larger than those found in Section \ref{sec:wavelength}, indicating that the PSF convolution introduces significantly stronger distortions than the effect of integrating over a finite spectral bandpass. Moreover, $\Delta$Area increases significantly with inclination, indicating that PSF convolution progressively alters the extracted SB curves for more edge-on viewing geometries. This behaviour can be understood physically: at large inclinations, PSF convolution enhances the mixing of emission originating from different locations and scattering angles along the line of sight, thereby distorting the extracted SB curves. This trend is further modulated by the disc opening angle. For small opening angles, the disc is geometrically thin, so most of the dust is concentrated close to the projected birth ring traced by the ellipses. Convolution then mixes light along this narrow structure, producing a noticeable change in the extracted SB. For larger opening angles, a larger fraction of the dust lies above or below the plane of the birth ring, outside of the elliptical apertures. Since the SB from these regions is not captured by the apertures in the first place, PSF convolution has less impact on the extracted SB.\\
\indent Since PSF convolution systematically degrades the fidelity of the extracted SB phase functions, particularly for highly inclined and vertically thin discs, all subsequent analyses are performed on unconvolved images. These therefore represent a best-case scenario for direct SPF extraction.

\subsection{Impact of dust size distribution and disc geometry}
\begin{figure*}[ht]
    \includegraphics[scale=0.24]{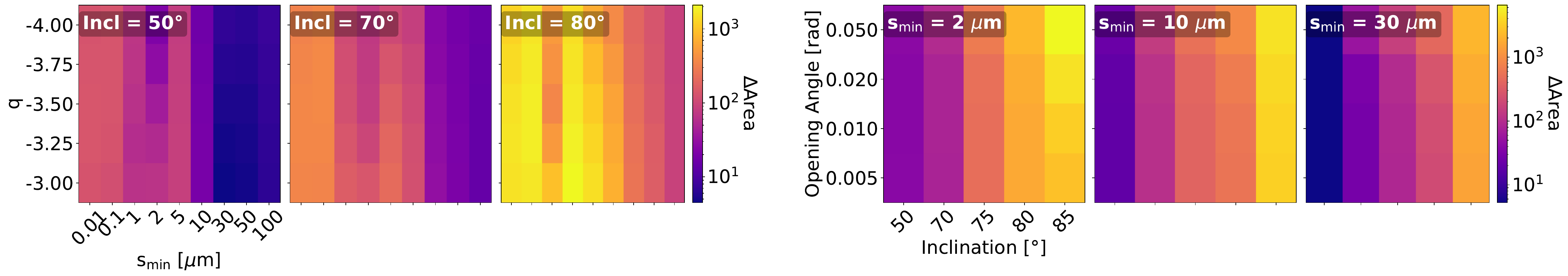}
    \caption{Heatmaps showing the integrated absolute difference ($\Delta$Area; Eq. \ref{eq:delta_area}) between the extracted SB and the intrinsic SPF for our synthetic disc models. The left panels show variations with size distribution slope, $q$, and minimum grain size, $s_\mathrm{min}$, at fixed geometry, for inclinations of 50$^{\circ}$, 70$^{\circ}$, and 80$^{\circ}$. The right panels display the variations with disc opening angle and inclination for $s_\mathrm{min}$ = 2, 10, and 30 micron.}
\label{fig:diff_SB_and_optool}
\end{figure*}
\noindent Having fixed the aperture definition, wavelength treatment, and excluded PSF convolution, we now examine the inherent limitations of SPF extraction arising from the dust size distribution and disc geometry alone. We again visualise this in a series of heatmaps, where we calculate the difference between the extracted SB and the input SPFs used in \texttt{bound} for various size distributions and geometries (Figure \ref{fig:diff_SB_and_optool}). Here, the input SPF refers to the phase function integrated over the full grain-size distribution between $s_\mathrm{min}$ and $s_\mathrm{max}$ with slope $q$, as computed with \texttt{optool}, and therefore represents the true scattering behaviour of the dust population. In the birth ring, the grain size distribution is expected to follow this same slope $q$; deviations due to radiation pressure primarily affect grains that populate regions outside the parent belt. Since our apertures are placed along the birth ring, the extracted SB is therefore expected to probe the same underlying size distribution as assumed in the input SPF, ensuring that the comparison is dynamically consistent.\\
\indent The $\Delta$Area heatmaps reveal several clear trends. In the $q$ vs $s_\mathrm{min}$ plots, the largest differences occur for small and intermediate grain sizes ($s_\mathrm{min} \lesssim 10$ micron). This behaviour is not primarily due to the intrinsic SPF becoming independent of grain size, but rather reflects how reliably the SPF can be recovered from the images. For small and intermediate grain sizes, the scattering phase functions exhibit strong forward-scattering peaks. These peaks are difficult to sample accurately with elliptical apertures, because a significant fraction of the forward-scattered light is concentrated into a narrow range of scattering angles and projected locations. Small mismatches between the true scattering geometry and the assumed mapping from sky position to scattering angle therefore lead to large discrepancies between the extracted SB and the input SPF, increasing $\Delta$Area. As $s_\mathrm{min}$ increases, the SPFs become less strongly forward-scattering (see Figure \ref{fig:gallery_bound}), and the extraction becomes correspondingly more stable and reliable, resulting in smaller $\Delta$Area values. Differences between various $q$ values are comparatively minor. Varying $q$ changes the relative weighting of grain sizes in the distribution, but does not fundamentally alter the geometric limitations of the extraction method. As a result, the dominant effect seen in the heatmaps is driven by how strongly forward-scattering grains contribute to the observed signal, rather than the precise slope of the size distribution.\\
\indent The inclination also plays a significant role. In the $s_\mathrm{min}$ vs $q$ heatmaps, $\Delta$Area increases with inclination because higher inclinations enhance line-of-sight projection effects, causing scattering from different vertical and radial locations to overlap along the projected ellipse. This overlap broadens the range of true scattering angles contributing to a given aperture, weakening the one-to-one correspondence assumed in the extraction procedure. A similar trend is visible in the opening angle vs inclination panels, where $\Delta$Area increases with both parameters, although the dependence is clearly dominated by the inclination. A larger opening angle spreads dust vertically, increasing the range of scattering angles present at a given projected position, while higher inclinations further amplify this mixing along the line of sight.\\
\indent Taken together, these results demonstrate that both the dust population and the disc geometry strongly influence the accuracy of SPF extraction from scattered-light images. In particular, the method is most reliable when the scattering phase function is relatively smooth (i.e. less dominated by a sharp forward peak) and when geometric projection effects are limited, corresponding to discs with larger characteristic grain sizes, small opening angles, and moderate inclinations.

\section{Henyey-Greenstein fitting}
\label{sec:HG}
\noindent Having quantified how disc geometry, projection effects, and methodological choices shape the extracted SPFs, we now examine how these effects propagate into commonly used parametric descriptions. In particular, Henyey-Greenstein phase functions (\citealt{Henyey_Greenstein1941}) are frequently used in debris disc studies to analyse the forward-scattering behaviour of dust grains. By fitting HG functions to the SB curves extracted from our synthetic images, we can directly test how reliably the inferred asymmetry parameters reflect the true dust scattering properties across different disc configurations and dust populations.
\subsection{Fitting method}
\label{sec:fitting_method}
\noindent In its single-component form, the HG phase function is defined as
\begin{equation}
    HG(g, \theta) = \frac{1}{4\pi} \frac{1 - g^{2}}{(1 - 2g\cos(\theta) + g^{2})^{3/2}} \ ,
\label{eq:HG}
\end{equation}
where $g$ is the asymmetry parameter, which quantifies the degree of anisotropy in the scattering. $g = 0$ corresponds to isotropic scattering, $0<g<1$ indicates forward scattering, and $-1<g<0$ indicates backward scattering. In observational studies, $g$ is often qualitatively interpreted as an indicator of grain size, with larger $g$ values commonly associated with more forward-scattering grains larger than the wavelength of the observations (see e.g. \citealt{Graham2007}; \citealt{Milli2017}; \citealt{Engler2020}; \citealt{Adam2021}). This practice implicitly assumes that variations in the extracted surface-brightness phase function can be meaningfully mapped onto changes in the underlying dust population through a single HG parameter.\\
\indent Here, we test this assumption explicitly by fitting the HG function to our extracted SPFs. However, a single HG function can only produce a forward or backward curve, whereas physically motivated SPFs computed from realistic dust models often show more complex shapes (e.g. strong forward-scattering peaks with back-scattering components). To better approximate these shapes, we adopt a two-component HG formulation, in which the total phase function is expressed as a weighted sum of two HG functions with different asymmetry parameters:
\begin{equation}
    HG2(g_\mathrm{1}, g_\mathrm{2}, w_\mathrm{1}, \theta) = w_\mathrm{1} \cdot HG(g_\mathrm{1}, \theta) + (1 - w_\mathrm{1}) \cdot HG(g_\mathrm{2}, \theta) \ ,
\end{equation}
where $g_\mathrm{1}$ usually describes the forward scattering and $g_\mathrm{2}$ the backward scattering. $w_\mathrm{1}$ is the scaling parameter, with $0\leq w_\mathrm{1}\leq1$.\\
\indent Throughout this work, we focus only on the $g_\mathrm{1}$ parameter, as this parameter is most commonly used in the literature to characterise the SPF. The 2-component HG parameters are fitted to the extracted, normalised SB curves using a Markov Chain Monte Carlo (MCMC) approach with the \texttt{emcee} package (\citealt{Foreman_Mackey2013}). To enable stable fitting and to prevent the results being driven by numerical noise, we assign approximate uncertainties to the normalised SB based on Poisson statistics. The MCMC sampler uses 48 walkers and 31,000 steps to ensure convergence and prevent the chains from becoming trapped in local likelihood minima. The first 3,000 steps are discarded as burn-in. Convergence was assessed by visually inspecting the parameter trace plots and by computing the integrated autocorrelation time for the fitted parameters. The total chain length corresponds to a median of $\approx$400 integrated autocorrelation times, indicating that the posterior distributions are well sampled and that the chains have converged. From the resulting posterior distribution, we report the median value of $g_\mathrm{1}$ as the best-fit forward-scattering parameter and derive uncertainties from the 16th and 84th percentiles.

\subsection{HG $g_\mathrm{1}$ variations with dust and geometry}
\label{sec:g1_vs_parameters}
\begin{figure*}[ht]
    \includegraphics[scale=0.24]{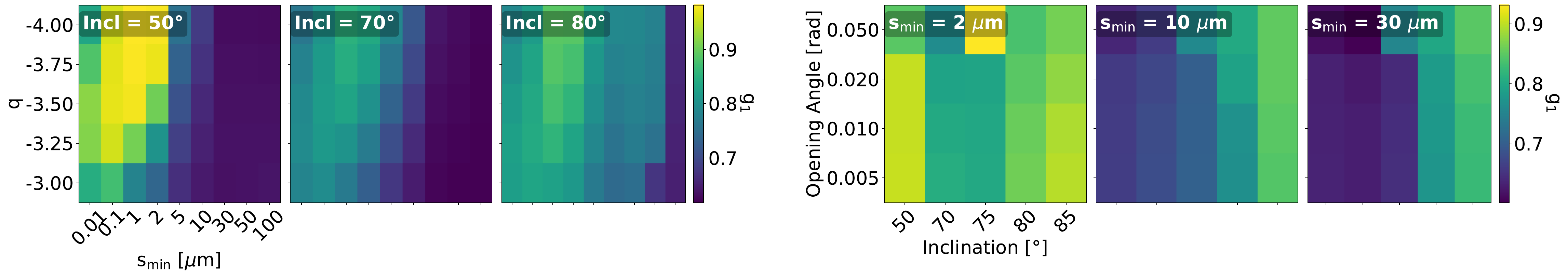}
    \caption{Heatmaps showing the dependence of the best-fit two-component HG asymmetry parameter $g_\mathrm{1}$ on dust properties and disc geometry. The left panels show $g_\mathrm{1}$ as a function of size distribution slope, $q$, and minimum grain size, $s_\mathrm{min}$, at fixed geometry, for inclinations of 50$^{\circ}$, 70$^{\circ}$, and 80$^{\circ}$. The right panels display $g_\mathrm{1}$ as a function of disc opening angle and inclination for $s_\mathrm{min}$ = 2, 10, and 30 micron.}
\label{fig:g1_matrix}
\end{figure*}
\noindent To visualise how the inferred $g_\mathrm{1}$ depends on the dust properties and viewing geometry, we again generate a series of heatmaps, shown in Figure \ref{fig:g1_matrix}. The first set of heatmaps on the left-hand side of the image shows $g_\mathrm{1}$ as a function of $q$ and $s_\mathrm{min}$ for the three fixed inclinations, while the second set displays $g_\mathrm{1}$ as a function of opening angle and inclination for the three fixed dust models. Each pixel is colour-coded by the best-fit $g_\mathrm{1}$ value. Similarly, we construct heatmaps of the uncertainty on $g_\mathrm{1}$, presented in Figure \ref{fig:g1_std_matrix}.\\
\indent The heatmaps reveal several clear trends. The largest $g_\mathrm{1}$ values occur for the smallest minimum grain sizes ($s_\mathrm{min} \lesssim 10$ $\upmu$m), with $g_\mathrm{1}$ decreasing steadily as $s_\mathrm{min}$ increases. This indicates that the inferred forward-scattering behaviour is strongest for grains with sizes similar to the wavelength in our models. This behaviour is consistent with the qualitative trends already evident in the synthetic disc images (Figure \ref{fig:gallery_bound}), where the strongest brightness asymmetries are observed for models with grain sizes comparable to the observing wavelength. Interestingly, different combinations of $q$ and $s_\mathrm{min}$ often yield very similar $g_\mathrm{1}$ values, demonstrating that there is no unique one-to-one mapping from the HG asymmetry parameter to the underlying dust properties.\\
\indent Geometry also plays a significant role, although the dependence on inclination is not uniform across parameter space. For models with small minimum grain sizes ($s_\mathrm{min} \lesssim 5$ $\upmu$m), $g_\mathrm{1}$ tends to decrease with increasing inclination. In contrast, for larger minimum grain sizes, $g_\mathrm{1}$ increases with inclination. This change of behaviour reflects a shift in regime: at higher inclinations, the observable scattering-angle range extends closer to the forward-scattering peak, allowing it to be better probed for grains that retain a pronounced forward lobe within the accessible angular range. All these trends largely reflect the influence of the SB extraction method. Because the elliptical apertures primarily sample the projected birth ring, the measured surface brightness does not capture the full distribution of scattered light.\\
\begin{figure*}[ht]
    \includegraphics[scale=0.25]{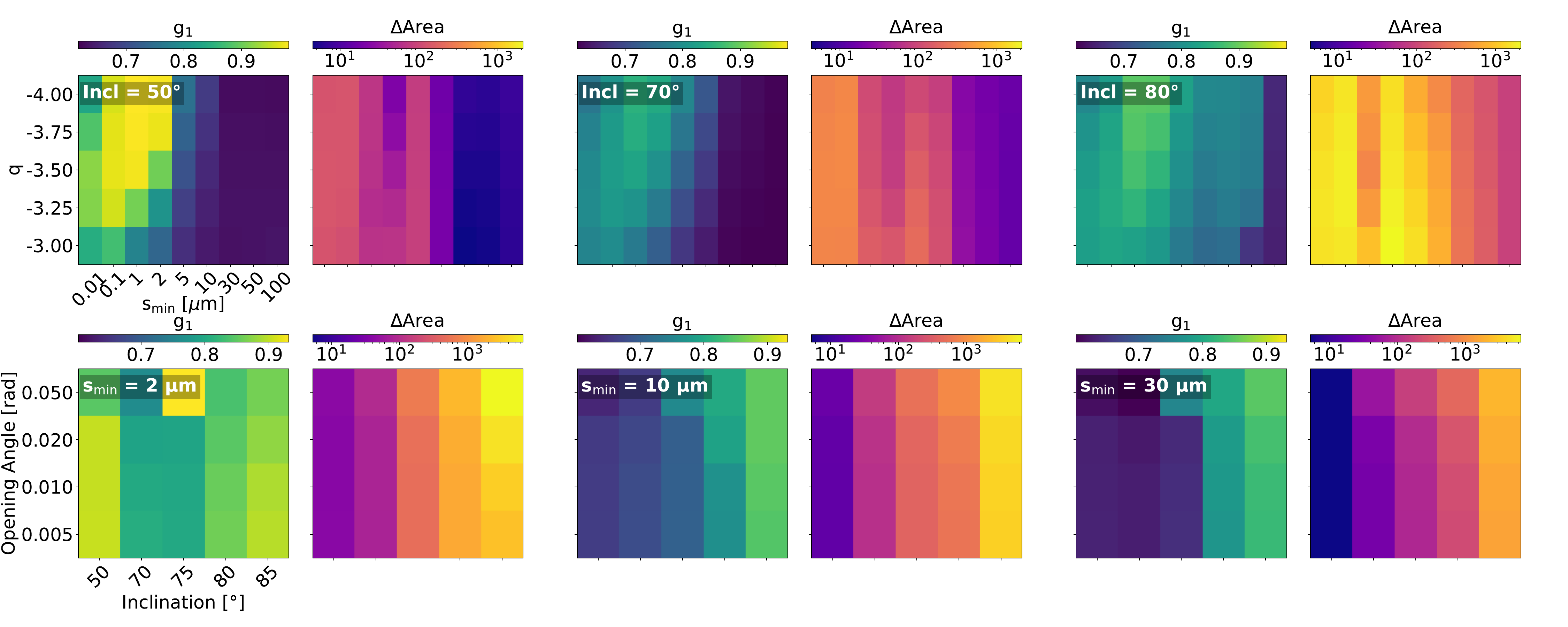}
    \caption{Heatmaps comparing the best-fit two-component HG asymmetry parameter $g_\mathrm{1}$ (left panels, same as Fig. \ref{fig:g1_matrix}) with the integrated absolute difference $\Delta$Area between the extracted surface brightness and the intrinsic scattering phase function (right panels, same as Fig. \ref{fig:diff_SB_and_optool}). The top rows show variations with size distribution slope, $q$, and minimum grain size, $s_\mathrm{min}$, at fixed geometry, for inclinations of 50$^{\circ}$, 70$^{\circ}$, and 80$^{\circ}$. The bottom row shows variations with disc opening angle and inclination for $s_\mathrm{min}$ = 2, 10, and 30 micron.}
\label{fig:comparing_g1_SB}
\end{figure*}
\indent This geometric and sampling bias becomes particularly apparent when comparing Figure \ref{fig:g1_matrix} to Figure \ref{fig:diff_SB_and_optool}, which we do in Figure \ref{fig:comparing_g1_SB}, showing the same heatmaps of $g_\mathrm{1}$ alongside the $\Delta$Area heatmaps between the extracted SB and the intrinsic SPF. The two sets of maps display similar patterns: parameter combinations that produce large $\Delta$Area values, most noticeably small minimum grain sizes ($s_\mathrm{min} \lesssim 10$ $\upmu$m), high inclinations, and large opening angles, also tend to yield larger $g_\mathrm{1}$ values. This indicates that the conditions under which the extracted SB deviates from the true SPF are also those in which the apparent forward-scattering behaviour is most enhanced. Consequently, both dust properties and disc geometry, together with the limitations of the sampling technique, determine how reliably the forward-scattering behaviour can be inferred from the images.

\section{Discussion}
\label{sec:discussion}
\subsection{Why recovering the intrinsic scattering phase function is inherently difficult}
\label{sec:discuss_why_recovering_spf_difficult}
\noindent Scattering phase functions inferred from scattered-light images are often interpreted as direct tracers of dust properties, in particular grain size. However, our results demonstrate that recovering the intrinsic SPF of debris discs from imaging observations is inherently challenging, even under idealised conditions where the disc geometry, the orbit of the disc ring, and the composition are known exactly from the \texttt{bound} simulations used to generate our synthetic images. These difficulties do not arise from a single limitation, but rather from a combination of disc geometry, observational viewing constraints, and methodological choices that fundamentally constrain what information can be extracted from scattered-light images.\\
\indent A primary limitation is the restricted range of scattering angles that can be probed in observations. For an inclined disc, the observable scattering-angle range is bounded by
\[
\qquad \qquad \theta_\mathrm{min} = 90^{\circ} - i, \qquad \qquad \theta_\mathrm{max} = 90^{\circ} + i,
\]
such that the smallest scattering angles remain entirely unobserved, even for highly inclined systems. This limitation is particularly important because the most diagnostic part of the SPF (the strong forward scattering peak), occurs at very small angles (see Figure 3 of \citealt{Mulders2013}). As a result, the observed surface brightness samples only a restricted portion of the underlying 
SPF, which is often comparatively flat.\\
\indent This effect provides a natural explanation for the counter-intuitive trends seen in the synthetic images presented in Section \ref{sec:synthetic_images}. As the minimum grain size increases, the phase function becomes increasingly forward scattering, but the forward peak shifts toward smaller angles, that fall outside the observable range. The remaining portion of the SPF that is sampled in the images therefore appears progressively flatter, causing discs dominated by large grains to appear more isotropic than expected. By contrast, grains with sizes comparable to the observing wavelength exhibit broader forward-scattering behaviour that extends to larger scattering angles, allowing a larger fraction of the anisotropic signal to fall within the observable angular range (see also \citealt{Mulders2013}). This effect alone can already produce apparent scattering behaviours that differ substantially from the true dust properties.\\
\indent Additionally, scattered-light observations provide a two-dimensional projection of a fundamentally three-dimensional structure. Even for geometrically thin discs, a given sky-projected location generally receives contribution from dust grains at multiple physical locations along the line of sight, each associated with a true scattering angle. This geometric mixing becomes increasingly severe for highly-inclined discs and for discs with larger opening angles, where vertical and radial extent further broaden the range of contributing scattering angles. Consequently, the surface brightness measured at a given position does not correspond to a unique scattering angle but instead represents a weighted average over multiple angles, such that the extracted SB curve is already a geometrically smeared version of the intrinsic SPF (e.g. \citealt{Stark2014}).\\
\indent The commonly used SPF-extraction method introduces additional uncertainties. This technique relies on sampling the disc with apertures placed along the projected ellipse that is assumed to trace the birth ring. While this approach is physically motivated and widely used, the exact choice of aperture size, spacing, and the shape is not uniquely defined. As we have shown in Section \ref{sec:ellipses}, reasonable variations in these choices can lead to measurably different extracted SB curves, even when applied to the same underlying image. This highlights that SPF extraction is not a strictly well-posed inverse problem: multiple, equally justifiable methodological choices can yield different apparent phase functions.\\
\indent Finally, real observations are unavoidably affected by instrumental point spread functions. PSF convolution mixes light from neighbouring pixels, further blending signals associated with different scattering angles and smoothing sharp features in the surface brightness distribution. For inclined and vertically thin discs in particular, this effect can significantly distort the extracted SB, thereby amplifying the geometric mixing already present in the unconvolved images.\\
\indent Taken together, these effects imply that discrepancies between inferred and intrinsic SPFs are not merely the result of imperfect data or suboptimal analysis techniques, but instead reflect fundamental observational and geometric limitations. Even in the absence of noise and with perfect knowledge of the disc architecture (e.g. inclination and PA), the true scattering phase function cannot, in general, be uniquely recovered from scattered-light images using direct extraction techniques based on surface brightness measurements alone. Forward-modelling approaches provide a more robust framework by explicitly incorporating the full three-dimensional dust distribution and its projection onto the sky, although they remain subject to their own assumptions and degeneracies. For example, such models often assume a single, spatially invariant phase function throughout the disc (e.g. in \texttt{DDiT}, \citealt{Olofsson2020}), applying equally to the birth ring and halo, despite the expectation that dust properties vary with location. These considerations have important implications for how SPFs and derived quantities, such as HG asymmetry parameters, should be interpreted physically in terms of dust properties.

\subsection{Implications for interpreting HG parameters}
\label{sec:discuss_implic_hg_params}
\noindent Our results have direct implications for how Henyey-Greenstein asymmetry parameters inferred from scattered-light observations should be interpreted. In observational studies, the asymmetry parameter $g$ is often treated as a proxy for grain size or overall scattering anisotropy, implicitly assuming that the extracted SB phase function provides a faithful representation of the underlying dust scattering behaviour. Our analysis shows that this assumption is generally not valid.\\
\indent We find that the inferred $g_\mathrm{1}$ is strongly influenced not only by the underlying dust properties, but also by disc geometry, viewing inclination, and the surface-brightness extraction method. In particular, parameter combinations that produce the largest deviations between the extracted SB and the intrinsic SPF also yield the largest apparent forward-scattering parameters. As a consequence, similar $g_\mathrm{1}$ values may correspond to very different physical dust populations, and conversely, inherently different SPFs may produce similar $g_\mathrm{1}$ values once projected onto the sky and analysed with the standard extraction technique. As a result, the HG asymmetry parameter should not be interpreted as a direct or unique diagnostic of grain size or composition, but rather as an effective, observation-dependent quantity that encodes both dust scattering physics and the geometric and methodological effects inherent to scattered-light imaging.\\
\indent An additional complication is that the apparent degree of forward scattering visible in scattered-light images does not necessarily vary monotonically with the true anisotropy of the dust grains. As shown by the synthetic images and quantitative analysis in this work, discs dominated by larger grains do not always appear more forward scattering than those dominated by smaller grains, because the strongest forward-scattering peak occurs at angles that are not accessible observationally. Consequently, the ordering of systems in terms of apparent anisotropy, and thus in terms of inferred $g_\mathrm{1}$, may not reflect the true ordering of intrinsic SPFs. This further limits the extent to which HG parameters can be interpreted in terms of grain size trends across different systems.\\
\indent These findings do not invalidate the use of HG functions as convenient empirical descriptions of scattered-light observations. However, they highlight that meaningful physical interpretations of HG parameters require careful consideration of geometric effects and observational limitations, ideally supported by forward-modelling approaches that explicitly account for projection, PSF convolution, and the restricted range of accessible scattering angles. While multi-component HG representations provide additional flexibility in fitting complex SB curves, they remain simplified parametrisations and do not bypass the fundamental observational and geometric limitations discussed before.

\subsection{Implications for low-mass stars}
\label{sec:discussion_implic_low_mass_stars}
\noindent The effects discussed above are not expected to impact all debris discs equally. In particular, for debris discs around low-mass stars, radiation pressure is weak and the classical radiation blow-out size is either very small or entirely absent, allowing grains with sizes well below a micron to remain gravitationally bound to the system (e.g. \citealt{Strubbe_Chiang2006}; \citealt{Pawellek_Krivov2015}). While such small minimum grain sizes may appear extreme in the context of debris discs around solar-type or earlier-type stars, they are physically plausible for discs around M dwarfs. Observationally, only a handful of discs have been detected around such low-mass stars to date (\citealt{Kalas2004}; \citealt{Low2005}; \citealt{Kennedy2013}; \citealt{Choquet2016}; \citealt{Sissa2018}; \citealt{Palatnick2025}).\\
\indent Our modelling shows that precisely these small $s_\mathrm{min}$ regimes are those in which the discrepancies between the extracted surface brightness and the intrinsic SPF are largest (see Figure \ref{fig:diff_SB_and_optool}), and where the apparent scattering behaviour seen in images departs most strongly from the underlying grain scattering properties. In these cases, geometric mixing, the limited range of accessible scattering angles, and details of the extraction procedure combine to produce large variations in the recovered phase functions and fitted HG parameters. As a result, the inferred HG asymmetry parameter is least constrained in these regimes, rather than most informative. This implies that debris discs around low-mass stars are particularly susceptible to misinterpretation when analysed using standard SPF-extraction techniques and HG fitting. In this sense, low-mass-star discs represent a "worst-case" scenario for the direct physical interpretation of HG parameters derived from scattered-light images.\\
\indent By contrast, discs around more luminous stars are expected to have larger minimum grain sizes set by the radiation pressure blow-out. In these systems, our results indicate that the extracted SB curves more closely resemble the true SPF, and the inferred HG parameters are correspondingly more stable. Nevertheless, even in this regime, the asymmetry parameter cannot be uniquely mapped to a specific grain size or dust population. As shown by \citet{Mulders2013} and by our work, similar HG asymmetry parameters can arise from a wide range of minimum grain sizes, underscoring that the degeneracy between scattering behaviour and grain size persists even when geometric biases are reduced.

\subsection{Implications for polarised scattered-light images}
\label{sec:discussion_polarisation}
\noindent Although our analysis focuses on total-intensity scattered-light images, many of the fundamental limitations identified in this work also apply to polarised scattered-light observations. Polarimetric images are subject to the same projection effects, line-of-sight mixing, PSF convolution, and aperture-dependent extraction biases that affect the total intensity. However, an important difference arises from the shape of polarised SPFs. In contrast to total intensity, polarised SPFs typically do not exhibit a strong forward-scattering peak; instead they generally peak near scattering angles of $\sim90^{\circ}$. As a result, the limited accessibility of small scattering angles, which strongly impacts total-intensity analyses, is less problematic in polarimetric studies. However, these more complex SPF shapes cannot be captured by a single asymmetry parameter (e.g. \citealt{Milli2019}; \citealt{Engler2020}). Therefore, while polarised observations are generally less sensitive to the specific bias introduced by missing small scattering angles, the broader methodological caveats identified in this work remain relevant.

\section{Conclusions}
\label{sec:conclusion}
\noindent Scattered-light images of debris discs are often interpreted in terms of dust properties, with scattering phase functions and Henyey-Greenstein asymmetry parameters commonly used as a proxy of grain size and scattering anisotropy. However, the brightness distribution observed in a scattered-light image does not necessarily trace the intrinsic scattering phase function. Projection effects, line-of-sight mixing, and the limited range of accessible scattering angles all shape the measured signal, and can lead to trends that appear counter-intuitive when interpreted in purely physical terms. In this work, we have performed, for the first time, a systematic test of the reliability of scattering phase function determinations using a physically motivated forward-modelling framework that combines dust dynamics, realistic optical properties, and ray-tracing imaging to generate synthetic observations. We used this framework to quantify how observational and geometrical effects influence the SPFS and HG parameters that are recovered using standard analysis techniques, and to assess how reliably these measurements can constrain dust properties. Our main conclusions are:
\begin{itemize}
    \item The apparent scattering behaviour seen in scattered-light images does not directly reflect the intrinsic dust scattering phase function. In particular, strongly forward-scattering grains may not appear as such in the observable angular range.
    \item The accessible scattering-angle range plays a central role in shaping the recovered phase functions. Because the smallest scattering angles, where the forward-scattering peak of large grains occur, are generally unobservable, the extracted surface brightness samples only a restricted and often relatively flat portion of the true SPF. As a result, the apparent scattering behaviour does not vary monotonically with grain size: discs with very small grains can appear more anisotropic than expected, discs with grains sizes comparable to the observing wavelength may appear more strongly forward-scattering within the observable angular range, and discs with grains much larger than the wavelength may appear comparatively isotropic.
    \item Standard SPF-extraction techniques introduce additional systematic uncertainties that are difficult to mitigate in practice. The inferred SPF depends on methodological choices such as aperture placement, aperture size, and the assumed disc geometry, as well as observational effects such as PSF convolution.
    \item Henyey-Greenstein asymmetry parameters derived from scattered-light images should be interpreted as effective, observation-dependent quantities rather than direct tracers of dust properties such as grain size. The fitted $g_\mathrm{1}$ values depend not only on the physical scattering behaviour of the grains, but also on disc geometry and SPF-extraction biases. As a result, there is no one-to-one correspondence between $g_\mathrm{1}$ and grain size or scattering anisotropy.
    \item The discrepancies between the true and recovered scattering behaviour are largest in regimes with very small minimum grain sizes. These conditions are expected to occur in debris discs around low-mass stars, where radiation pressure is weak. Such systems therefore represent particularly challenging cases for interpreting the SPFs in terms of dust properties.
    \item Robust interpretation of scattered-light observations requires forward-modelling approaches that explicitly account for projection effects, limited scattering-angle coverage, and observational biases.
\end{itemize}

\begin{acknowledgements}
This research has made use of the SVO Filter Profile Service "Carlos Rodrigo", funded by MCIN/AEI/10.13039/501100011033/ through grant PID2023-146210NB-I00. We thank the referee for a careful reading of the manuscript and for constructive comments that helped improve the clarity of the paper.
\end{acknowledgements}

\bibliographystyle{aa.bst}
\bibliography{ReferencesPaper.bib}

\begin{appendix}
\section{Uncertainties on HG $g_\mathrm{1}$}
\label{appendix}
\begin{center}
\begin{minipage}{\textwidth}
    \centering
    \includegraphics[scale=0.24]{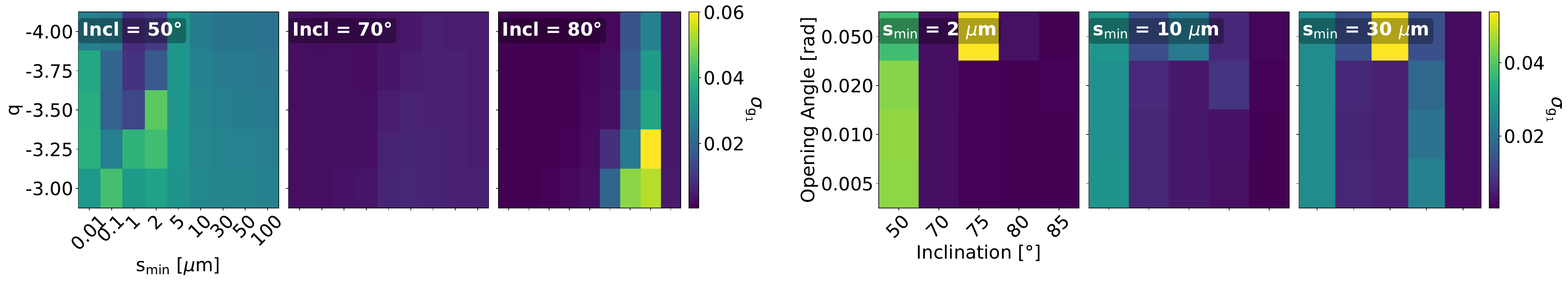}
    \captionof{figure}{Heatmaps showing the uncertainty on the best-fit two-component HG asymmetry parameter $g_\mathrm{1}$ for the same grid as in Figure \ref{fig:g1_matrix}.}
    \label{fig:g1_std_matrix}
\end{minipage}
\end{center}
\vspace{1em}
\noindent Figure \ref{fig:g1_std_matrix} shows the uncertainties on the fitted $g_\mathrm{1}$ values, derived from the 16th and 84th percentiles of the MCMC posteriors. On average, the uncertainties are small across the explored parameter space. Slightly larger uncertainties are seen for low-inclination discs (50$^{\circ}$) with large opening angles, likely reflecting weaker constraints on the forward-scattering component in these configurations. No strong systematic trends are apparent.

\end{appendix}

\end{document}